# Superinjection in diamond homojunction P-I-N

I.A. Khramtsov, and D.Yu. Fedyanin[a)]

*Laboratory of Nanooptics and Plasmonics, Moscow Institute of Physics and Technology, 141700 Dolgoprudny, Russian Federation*

a) dmitry.fedyanin@phystech.edu

Diamond and many newly emerged semiconductor materials show outstanding optical and magnetic properties. However, they cannot be as efficiently doped as silicon or gallium arsenide, which limits their practical applicability. Here, we report a superinjection effect in diamond p-i-n diodes, which gives the possibility to inject orders of magnitude more electrons into the i-region of the diode than the doping of the n-type injection layer allows. Moreover, we show that the efficiency of electron injection can be further improved using an i-p grating implemented in the i-region. The predicted superinjection effect enables to overcome fundamental limitations related to the high activation energy of donors in diamond and gives the possibility to design high-performance devices.

The operating principle of many semiconductor devices is based on the possibility to create a high density of nonequilibrium carriers under a bias voltage. These carriers can further recombine or change the properties of the semiconductor, which can, for example, be exploited for light modulation.[1] Light emitting devices, ranging from LEDs and lasers to single-photon sources, completely rely on this process.[2–5] In this regard, diamond, which was recently emerged as a material not only for high power electronics but also for room and high-temperature light emitting devices,[5–11] is undoubtedly inferior to most semiconductors like gallium arsenide. The extremely high activation energy of donors (~0.6 eV)[12] along with the non-zero donor compensation by acceptor-type defects limit the maximum density of free electrons in n-type diamond to only $10^{10} - 10^{11}$ cm$^{-3}$ at room temperature,[12,13] regardless the concentration of donor atoms. At the same time, the maximum density of free electrons in the active region of homojunction optoelectronic devices is typically limited by the electron density in the n-type injection layer,[14] which does not give diamond-based devices the possibility to compete with their more electron-rich semiconductor counterparts.

There are two possible strategies to overcome the electron density limitation. The first is extensive: by doping the n-type injection layer heavily with phosphorus, one creates a high density of hopping electrons.[15,16] These hopping carriers cannot be as efficiently used for recombination processes as electrons in the conduction band of diamond, but one can try to inject them into the undoped i-region region of the p-i-n structure and partially convert to conduction-band electrons. The second strategy relies on the possibility to create a smart structure that can inject a higher electron density than the n-type injection layer contains. Such an intensive approach is used in heterostructure LEDs and lasers, where due to the potential barriers at the heterojunction, it is possible to accumulate high densities of non-equilibrium electrons and holes that can be orders of magnitude higher than the free carrier densities in the n-type and p-type injection layers. This phenomenon known as superinjection[17,18] was claimed to be a unique property of heterostructures and is not met in semiconductor homojunctions.[19] This statement is correct for conventional semiconductor materials such as silicon or gallium arsenide. However, diamond is a unique material at the interface between insulators and semiconductors. This feature allows diamond diodes to operate in regimes, in which conventional semiconductor diodes burn out due to extremely high injection currents. In this work, using a rigorous numerical approach, we predict for the first time the superinjection effect in homojunction diamond diodes. We show that in the i-type region of the p-i-n diode, one can accumulate a density of free electrons that is nearly four orders of magnitude higher than the electron density in the n-type injection layer of the diode. Moreover, we show that the efficiency of electron injection can be further improved by implementing an i-p grating inside the i-region of the p-i-n diode. This approach gives the possibility to inject a high density of electrons into a large volume. The enhanced efficiency of electron injection can be used to greatly increase the brightness of single-photon sources and light emitting diodes based on diamond.

Figure 1(a) shows a schematic of the p-i-n diamond diode. The n-type region is doped with phosphorus at a concentration of $10^{18}$ cm$^{-3}$. The donor compensation ratio by acceptor-type defects is 10%, which is typical for n-type diamond samples[20,21] and provides a density of free electrons of $n_{eq} = 6 \times 10^{10}$ cm$^{-3}$. The concentration of acceptors (boron) in the p-type region is equal to $10^{18}$ cm$^{-3}$, and the acceptor compensation ratio is 1%[22]. The thickness of the i-region is selected to be 10 μm.[13] Such a thickness gives the possibility to observe the superinjection effect at relatively low injection currents. Other parameters used in the numerical simulations are listed in Table S1 in supplementary material. The p-i-n diode is simulated using the nextnano++ software (nextnano GmbH, Munich, Germany),[23] which is based on the self-consistent steady-state model (which comprises the Poisson equation for the electric field and carrier densities, semiconductor drift-diffusion equations for electrons and holes, and carrier continuity equations) and the finite difference method.

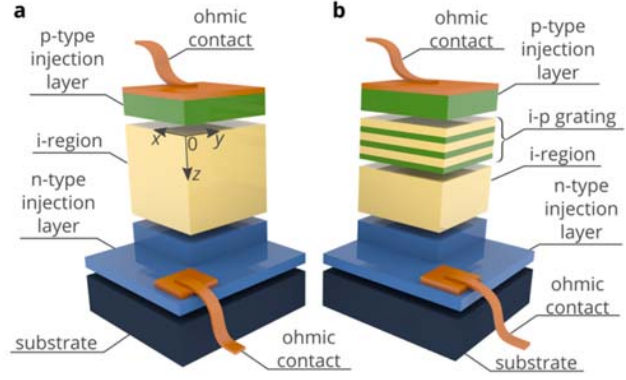

FIG. 1. Schematic illustrations of the p-i-n (panel a) and p-(i-p)$_N$-i-n (panel b) diamond diodes. $N$ is responsible for the number of periods of the i-p grating implemented in the i-region of the diode structure.

The results of the numerical simulations of the p-i-n diode are shown in Fig. 2 and 3. In equilibrium, there are no free carriers in the i-region. At forward bias voltages $V \gtrsim 4$ V, electrons and holes are injected from the n-type and p-type regions, respectively. At $V = 4.4$ V ($J = 0.5$ mAcm$^{-2}$), the density of electrons in the i-region is as high as $2 \times 10^9$ cm$^{-3}$ [Fig. 2(e)], but it is still lower than the electron density $n_{eq} = 6 \times 10^{10}$ cm$^{-3}$ in the n-region, which is typical for semiconductor diodes.[14,24] The grey line in Fig. 3 shows that at $V = 4.6$ V ($J = 0.1$ Acm$^{-2}$), the density of electrons at $z = 1.9$ μm exceeds $n_{eq} = 6 \times 10^{10}$ cm$^{-3}$. The electron density in the i-region continues to increase as the bias further increases, and the position of the maximum in the electron distribution shifts towards the p-i junction. Figures 2 and 3 demonstrate that at high bias voltages, the maximum density of electrons in the i-region is orders of magnitude higher than the electron density in the n-region. This is what is known as the superinjection effect, which was previously predicted and observed only in semiconductor heterostructure.[17,18] In the diamond p-i-n homojunction diode, the superinjection effect arises at voltages well above the diode turn-on voltage. Such superinjection conditions cannot be achieved in conventional semiconductor p-n or p-i-n homojunction diodes at feasible currents since the current density increases with the bias voltage orders of magnitude faster than in diamond diodes.

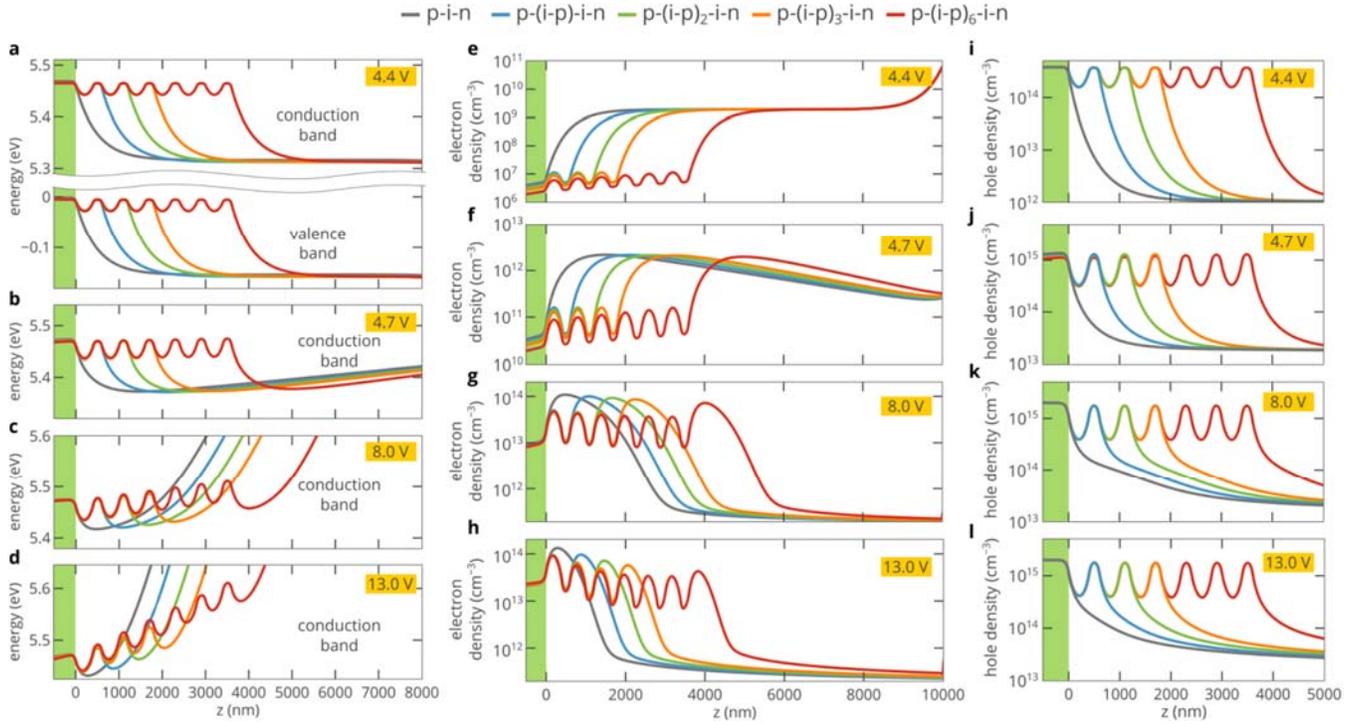

FIG. 2. (a) Energy band diagrams of the p-(i-p)$_N$-i-n diodes ($N$ = 0, 1, 2, 3, 6) at a forward bias voltage of 4.4 V. Hereinafter, the thicknesses of the i-type and p-type regions of the i-p grating are equal to 400 nm and 200 nm, respectively, the grating period is 600 nm. The green area indicates the p-type injection layer. (b-d) Relative position of the conduction band edges of the p-(i-p)$_N$-i-n diodes at bias voltages of 4.7 V (panel b), 8 V (panel c), and 13 V (panel d). (e-h) Electron density distribution in the p-(i-p)$_N$-i-n diodes at four different bias voltages. (i-l) Hole density distribution in the p-(i-p)$_N$-i-n diodes at four different bias voltages. The I-V characteristics of the p-(i-p)$_N$-i-n diodes can be found in supplementary material.

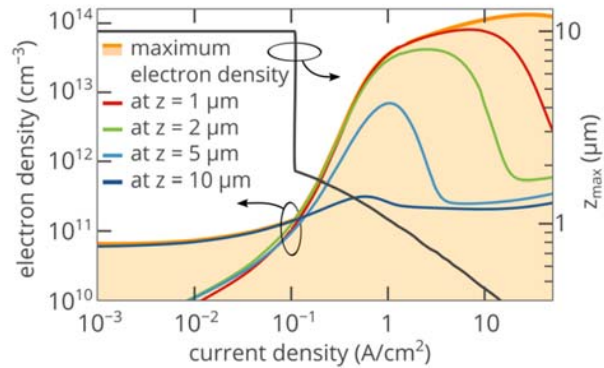

FIG. 3. Dependence of the electron density on the injection current at four different points in the i-region of the p-i-n diode shown in Fig. 1(a). The orange line shows the maximum electron density that can be created in the i-region of the diode, and the grey line shows the position of this maximum.

At bias voltages above $\sim 4.6$ V, the diamond p-i-n diode can be divided into two regions. In the first region (p-layer), the band bending (i.e., the electric field) is very weak compared with that in the second region (i- and n-layers), since the density of free carriers in the p-layer is much higher than in the i- and n-layers (see Fig. S2 in supplementary material). Therefore, the diffusion electron transport dominates in the p-layer, i.e., the electron current can be expressed as $J_n = qD_n \nabla n \approx qD_n n/L_n$, where $q$ is the elementary charge, $D_n$ is the electron diffusion coefficient and $L_n$ is the electron diffusion length. At the same time, the drift electron transport dominates in the i- and n-layers, i.e., $J_n = q\mu_n En$, where $E$ is the electric field and $\mu_n$ is the electron mobility. Since the recombination rate in the i-layer is relatively low due to the low density of defects, the electron currents at the p-i and i-n junctions should be of the same order of magnitude. At high bias voltages, the electric field in the n-layer is so high that the electron density in the p-layer near the p-i junction exceeds the electron density in the n-type layer. In addition, the electron density gradient $\nabla n = dn/dz$ is positive at the p-i junction ($z = 0$), while the electron density in the n-layer is lower than in the p-layer under this high voltage condition. Therefore, a transition region with a high electron density, which exceeds the electron densities in both the n-type and p-type layers, exists in the i-type layer near the p-i junction to ensure current continuity. The formation of this transition region is inevitably accompanied by the formation of a potential well (see Fig. 2(b-d)) since the local increase in the electron density is accompanied by the local decrease in $(E_c - F_n)$ and vice versa ($E_c$ is the conduction band edge and $F_n$ is the quasi-Fermi level for electrons). At high voltages, the electron density in the transition region can be as high as $10^{14}$ cm$^{-3}$ [Fig. 3], while the density of free electrons in the n-type injection layer is only $6 \times 10^{10}$ cm$^{-3}$. However, at high injection levels, the width of the region with a high electron density is as low as $300 - 500$ nm (see Fig. 2(g,h) and Fig. 3), while the remaining volume of the 10-μm-thick i-region is not used efficiently. Moreover, the width of the region with a high electron density only decreases as the bias voltage increases.

To improve the efficiency of electron injection, one could use a double heterostructure to create a wide potential well for electrons, as is done in III-V semiconductor light emitting devices.[2,3,17,18] However, in the case of diamond, we do not have compound semiconductors that are lattice matched to diamond and have a narrower bandgap perfectly aligned with respect to the conduction band of diamond.

We propose to use an i-p grating implemented in the i-region of the p-i-n diode [Fig. 1(b)]. By selectively doping the i-region, we can artificially control the band bending in the i-region near the p-i junction at high voltages and create potential wells for electrons [Fig. 2(c,d)], where it is possible to efficiently accumulate a high density of electrons at voltages above 7 V [Fig. 2(g,h)]. Figure 2(e-g) clearly shows that a single 600-nm-thick i-p insert (which consists of the 400-nm-thick i-type layer and the 200-nm-thick p-type layer) in the i-region of the p-i-n diode does not affect the electron distribution in the remaining i-region of the p-i-n diode at moderately high bias voltages. Therefore, this i-p insert additively contributes to the injection efficiency not deteriorating the superinjection effect in the remaining i-region. Remarkable is that even a thick i-p grating with six periods of 600 nm does not change the maximum electron density [Fig. 2(f,g)] and only slightly reduces the width of the area with a high electron density in the remaining i-region due to the decreased thickness of the remaining i-region and increased influence of the i-n junction on the electron distribution in the i-region.[13] At the same time, electrons are efficiently injected into the i-p grating creating a wide region with a high electron density [Fig. 2(f,g)], which can be efficiently exploited in the design of light emitting devices. At bias voltages above 11 V [Fig. 2(d,h,l)], the influence of the i-p grating on the remaining i-region is stronger. However, we should note that such regimes are less favorable for light emitting devices due to the much higher currents and slightly lower electron densities [Fig. 2(g)]. Nevertheless, the i-p grating can significantly improve the diode characteristics in these regimes. Figure 2(g) shows that at bias voltages above 11 V, the six-period grating slightly decreases the maximum density of electrons in the i-region. However, the total number of injected electrons between the n-type and p-type injection layers of the diode is 25% higher than in the p-i-n diode without the grating and the width of the region with a high electron density is 5 times larger.

It is important that at moderately high bias voltages ($V < 11$ V), the electron density in each cell of the i-p grating is absolutely the same in spite of the large length of the i-p grating and the relatively strong electric field in the i-region [Fig. 2(f,g)]. Color centers created in the i-type inserts of the i-p grating would demonstrate absolutely the same emission rates and other properties [Fig. 4], which can be beneficial for the design of reproducible electrically pumped single-photon sources or light emitting diodes based on color centers in diamond.

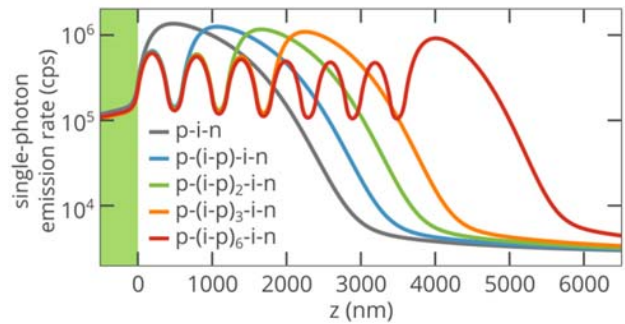

FIG. 4. Single-photon electroluminescence rate of the NV center versus its position in the p-(i-p)$_N$-i-n diodes at a bias voltage of 8 V. The electroluminescence rate is calculated at room temperature using the theory from Refs. 25,26, the quantum efficiency of the NV center is 78% and the radiative lifetime of the excited state is 17 ns.[7,25,27]

Another important advantage of the implementation of the i-p grating is the greatly improved hole injection efficiency, which is clearly seen in Fig. 2(i-l). The i-p grating gives the possibility to accumulate holes across the whole i-p grating. The maximum density of holes in the p-type inserts of the grating is as high as in the p-type injection layer, while the hole density in the i-type inserts is only four times lower. The improved hole injection efficiency can greatly increase the number of free excitons [Fig. 5], especially at high forward bias voltages [Fig. 5(b)], since the exciton density is determined equally by both the electron and hole densities.[28]

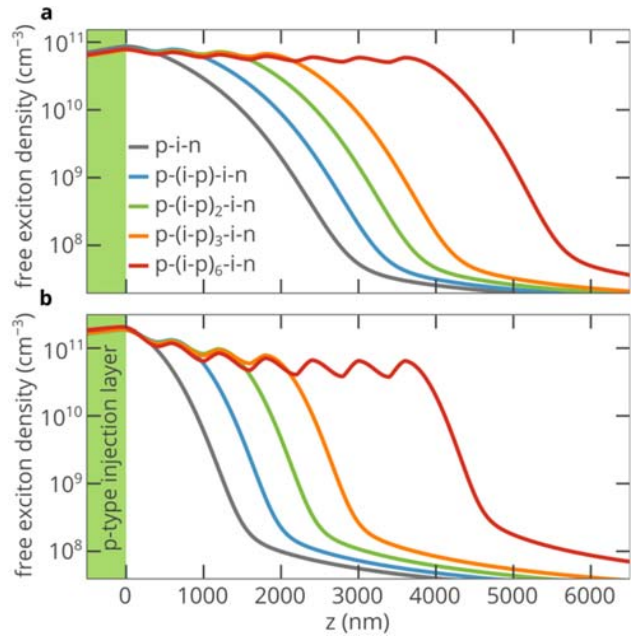

FIG. 5. Density of free excitons at 300 K calculated using the mass-action law[28,29] and the densities of electrons and holes obtained from the numerical simulations of the p-(i-p)$_N$-i-n diodes at bias voltages of 8 V (panel a) and 13 V (panel b).

In summary, we have numerically demonstrated that although the high activation energy of donors in diamond limits the density of free electron in the n-type injection layer to only $10^{10} - 10^{11}$ cm$^{-3}$, it is possible to inject much more carriers into the i-region of the p-i-n diamond diode. Moreover, we have shown that the electron injection efficiency can be further improved using an i-p grating implemented in the i-region near the p-i junction. Until recently, it was believed that the superinjection effect was a distinct feature of semiconductor heterostructures[17,19] and could not be observed in p-n and p-i-n homojunctions. However, we have shown that diamond-based diodes can operate in the regimes that are unreachable by conventional semiconductors, which allows to reach the superinjection conditions and design efficient electronic and optoelectronic devices. Such structures that exploit the superinjection effect open new opportunities in the design and engineering of high-performance optoelectronic devices based on diamond and other wide-bandgap semiconductors.

See supplementary material for the parameters used in the simulations, I-V characteristics and electric field distributions.

This work is supported by the Russian Science Foundation (17-79-20421).

# Supplementary material for
# Superinjection in diamond homojunction P-I-N diodes

## Igor A. Khramtsov and Dmitry Yu. Fedyanin*

*Laboratory of Nanooptics and Plasmonics, Moscow Institute of Physics and Technology, 141700 Dolgoprudny, Russian Federation*

*E-mail: dmitry.fedyanin@phystech.edu

Table S1. Parameters used in the numerical simulations.

| Parameter | Value |
|---|---|
| Energy band gap of diamond | 5.47 eV[1] |
| Dielectric constant of diamond | 5.7[1] |
| Density of donors in the n-type region | $10^{18}$ cm$^{-3}$ |
| Activation energy of donors | 0.57 eV[2] |
| Donor compensation ratio, $\eta_n$ | 10 % |
| Density of acceptors in the p-type region | $10^{18}$ cm$^{-3}$ |
| Activation energy of acceptors | 0.37 eV[3] |
| Acceptor compensation ratio, $\eta_p$ | 1 % |
| Background density of donors in the i-type region | $10^{14}$ cm$^{-3}$ |
| Background density of acceptors in the i-type region | $10^{14}$ cm$^{-3}$ |
| Electron and hole recombination lifetimes in the p-type region | 2 ns[a] |
| Electron and hole recombination lifetimes in the i-type region | 1.6 μs[a] |
| Electron and hole recombination lifetimes in the n-type region | 2 ns[a] |
| Electron mobility in the n-type and p-type regions | 740 cm$^2$/Vs[1,4] |
| Hole mobility in the n-type and p-type regions | 660 cm$^2$/Vs[1,4] |
| Electron mobility in the i-type region | 2500 cm$^2$/Vs[5] |
| Hole mobility in the i-type region | 1200 cm$^2$/Vs[5] |
| Longitudinal effective electron mass in the conduction band of | 1.56$m_0$[6] |
| diamond | |
| Transverse effective electron mass in the conduction band of diamond | $0.28m_0$[6] |
| Heavy-hole effective mass | $0.67m_0$[6] |
| Light-hole effective mass | $0.26m_0$[6] |
| Thickness of the n-type layer | 2 μm |
| Thickness of the p-type layer | 4 μm |
| Thickness of the i-type region | 10 μm |
| Thickness of the i-type layer of the i-p grating | 400 nm |
| Thickness of the p-type layer of the i-p grating | 200 nm |
| Mesh size | 1 - 50 nm[b] |
| Contacts type | ohmic[7] |

a) The recombination lifetimes for electrons and holes are calculated using the Scharfetter relation:[8]

$$\tau = \frac{\tau_0}{1 + \frac{N_D + N_A}{N_{ref}}}$$

where $N_D$ and $N_A$ are the concentrations of donors and acceptors, respectively, $N_{ref}$ is estimated to be[9–14] $10^{15}\,cm^{-3}$ and $\tau_0 = 2$ μs.[9] Despite that the Scharfetter relation may slightly overestimate the carrier recombination lifetimes in the i-region of the p-i-n diode, our numerical simulations show that the carrier recombination lifetimes in the i-region in the range from $10^{-8}$ s to $10^{-5}$ s almost do not affect the results of the numerical simulations under the superinjection conditions ($V \gtrsim 4.6$ V).

b) The mesh size was as low as 1 nm near the p-i and i-n junctions, and near the contacts. At the same time, in the bulk, the mesh size was up to 50 nm.

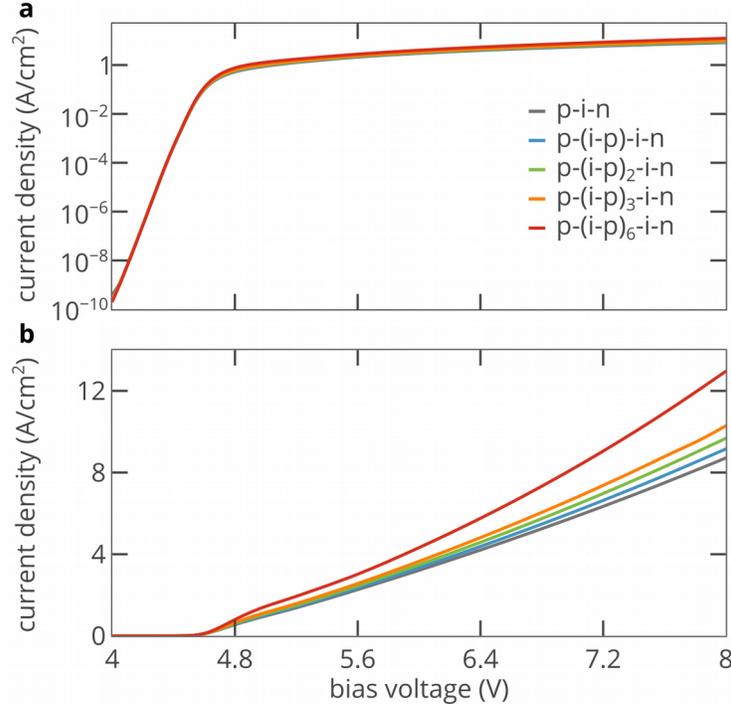

**Fig. S1.** Current-voltage characteristics of the p-(i-p)$_N$-i-n diodes ($N = 0, 1, 2, 3, 6$) in the logarithmic (a) and linear (b) scales.

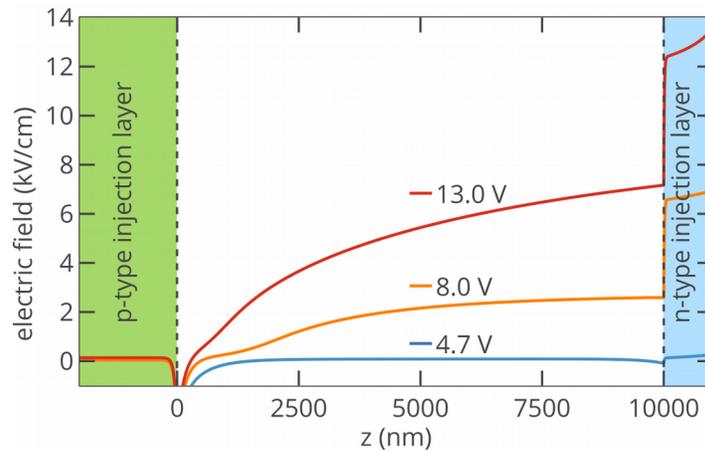

**Fig. S2.** Electric field distribution in the p-i-n diode at three different forward bias voltages.